\documentclass{PNAStwo}

\usepackage{graphicx}
\usepackage{PNAStwoF}
\usepackage{amssymb,amsfonts,amsmath}

\begin{document}

\conflictofinterest{The authors declare no conflict of interest}

\track{This paper was submitted directly to the PNAS office.}

\footcomment{Author contributions: J.W.S., D.A.F. designed the research; J.W.S. performed the research; J.W.S., D.A.F. wrote the paper}

\footcomment{Abbreviations: BR, Brownian Ratchet}

\title{Load fluctuations drive actin network growth}

\author{Joshua W. Shaevitz\affil{1}{Department of Integrative Biology}\thanks{Current address: Department of Physics and Lewis-Sigler Institute for Integrative Genomics, Princeton University Princeton, NJ 08544} \thanks{Email: shaevitz@princeton.edu} ,
Daniel A. Fletcher
\affil{2}{Department of Bioengineering, University of California, Berkeley, CA 94720, USA}\thanks{E-mail: fletch@berkeley.edu}}

\contributor{Submitted to Proceedings of the National Academy of Sciences of the United States of America}

\maketitle

\begin{article}

\begin{abstract}
The growth of actin filament networks is a fundamental biological process that drives a variety of cellular and intracellular motions. During motility, eukaryotic cells and intracellular pathogens are propelled by actin networks organized by nucleation-promoting factors, which trigger the formation of nascent filaments off the side of existing filaments in the network. A Brownian ratchet (BR) mechanism has been proposed to couple actin polymerization to cellular movements, whereby thermal motions are rectified by the addition of actin monomers at the end of growing filaments. Here, by following actin--propelled microspheres using three--dimensional laser tracking, we find that beads adhered to the growing network move via an object--fluctuating BR. Velocity varies with the amplitude of thermal fluctuation and inversely with viscosity as predicted for a BR. In addition, motion is saltatory with a broad distribution of step sizes that is correlated in time. These data point to a model in which thermal fluctuations of the microsphere or entire actin network, and not individual filaments, govern motility. This conclusion is supported by Monte Carlo simulations of an adhesion--based BR and suggests an important role for membrane tension in the control of actin--based cellular protrusions.
\end{abstract}


\dropcap{S}everal, non--mutually exclusive models have been proposed to link actin network growth to motion production, including those based on elastic gel compression, autocatalytic filament nucleation and the rupture of network--object linkages\cite{JWS:Mog2006}. The elastic-gel model accounts for the role of surface curvature on the movement of actin-propelled beads and bacteria, and it predicts an asymmetric distribution of stresses that results in propulsion through a squeezing mechanism\cite{JWS:Ger2000}. Autocatalytic branching models consider the effect of load variations on an actin network pushing a planar surface and predict a flat force-velocity relationship due to a filament number that varies in proportion with load\cite{JWS:Car2003}. Finally, adhesion-based models postulate that the rupture of chemical bonds between the moving object and actin network leads to a local relaxation of network stress and produces motion\cite{JWS:Mog2003,JWS:Soo2005}. While these models address different aspects of actin network organization and motility, each relies on the same underlying means of generating force; the Brownian ratchet. This model has become generally accepted as describing force generation by polymerizing actin networks, even though it has never been verified experimentally\cite{JWS:Pol2003}.

Peskin \textit{et al.} first proposed that actin polymerization could act as a ratchet, whereby the free energy of monomer addition at the ends of filaments is used to rectify the thermal fluctuations of a moving object\cite{JWS:Pes93}. It has also been shown theoretically that the rectification of the bending fluctuations of individual filaments within the network will generate a propulsive force, even when larger--scale fluctuations are negligible\cite{JWS:Mog96} (for simplicity, these two classes of BR models are referred to as ``object--fluctuating'' and ``filament--fluctuating'' herein). Because both object--fluctuating and filament--fluctuating models appear able to drive motility, it remains unclear which class of models might dominate.

A key feature of BR models is the dependence of velocity on the diffusion coefficient, $D$, of the relevant fluctuating element. When thermal motions of a moving object are instantly rectified over a distance scale $\delta$, forward movement proceeds at a velocity $v=2D/\delta$ (ref \cite{JWS:Pes93}). However, the assumption that thermal fluctuations are instantly rectified is not accurate in many situations\cite{JWS:Pes93} and slow actin monomer addition kinetics can lead to a lack of dependence of the velocity on $D$. In general, the effect of the diffusion coefficient on velocity depends on a detailed comparison of multiple timescales and can be nontrivial in most realistic models of actin--based motility which involve a dynamic number of filaments transiently attached to the moving object.

Data to support either object--fluctuating or filament--fluctuating BR models have been inconclusive. Several groups have attempted to probe changes in the object diffusion coefficient either by using beads of different radii or by changing the surrounding fluid viscosity. Bernheim-Groswasser \textit{et al.} found that velocity was roughly proportional to the inverse of the bead radius and thus the diffusion coefficient\cite{JWS:Ber2002}, although two other studies produced seemingly contradictory results\cite{JWS:Wie2003,JWS:Cam2004}. While changes in bead radius do affect the diffusion coefficient, they also change the size and potentially the topology of the polymerizing actin network, and this may result in bead velocity changes that are unrelated to the BR mechanism. In other experiments, live bacteria and beads were found to move slower when methylcellulose, an agent commonly used to increase viscosity and dampen thermal fluctuations, was added to the solution\cite{JWS:Cam2004,JWS:McG2003}.  However, a nonviscous form of methylcellulose was also found to slow moving beads and increase the total amount of actin within the network\cite{JWS:Wie2003}, making interpretation of these results difficult. Additionally, polymer solutions of methylcellulose are highly nonNewtonian and have complex rheological responses in time and space whose effects on the Brownian Ratchet are unclear\cite{Ber1979,Ama2001}.

\section{Results and Discussion}
To probe the effect of changes in the object diffusion coefficient alone while avoiding the above complications, we monitored the motion of actin-propelled beads near glass surfaces. The apparent diffusion coefficient of a bead changes as the bead approaches the surface, whereas the diffusion of proteins, chemical concentrations and enzyme kinetics remain unaffected. A typical 3D trajectory of actin--driven microsphere movement with accompanying images of fluorescent actin is shown in Fig.~1. We performed experiments in thin, $2.5 \pm 0.3$ $\mu$m thick chambers using spacer beads to separate a glass coverslip and slide (errors are listed as standard deviation unless otherwise stated).

The effective solution viscosity close to a flat wall is increased by proximity to the solid--liquid boundary. The magnitude of this increase depends on the ratio of the object radius to the object--glass separation. A one-micron diameter sphere located one micron from a flat surface experiences a 39\% increase in apparent viscosity, whereas a 5--nm sized protein at the same position feels only a 0.3\% increase. This hydrodynamic effect is enhanced in our chambers by the presence of two close--by interfaces, and the viscosity is given by\cite{JWS:Lin2000,JWS:Duf2001}
\begin{equation}
\eta \left(r,h\right)=\frac{\eta_\infty}{1-\frac{9r}{16}\left(\frac{1}{h}+\frac{1}{L-h}\right)}
\label{ViscEq}
\end{equation}
where $\eta\left(r,h\right)$ is the viscosity for a separation $h$ between the center of the bead and the closest glass surface, $\eta_\infty$ is the viscosity in an unbounded fluid, $r$ is the sphere radius ($396 \pm 12$~nm in our case), and $L$ is the separation of the two walls.

We recorded the motion of 13 actin-propelled beads for 30 minutes each or until the feedback tracking lost the bead, yielding run lengths from 8 to 173~$\mu$m. Bead velocities in these chambers averaged 58~$\pm$~38 nm/s. We measured the viscosity of our extract solutions using both a 1~mm diameter falling ball viscometer and by tracking the thermal motion of optically trapped beads. The average value of $\eta_\infty=2.8 \pm 0.2\times 10^{-3}$~Pa$\cdot$s is approximately three times that of water, similar to previously reported measurements\cite{JWS:Cam2004}.

We calculated the velocity as a function of the distance to the nearest glass surface and found that a viscous interaction with the chamber surfaces reduced the velocity of moving beads (Fig.~2a). Because of the wide variance in velocities from bead to bead, we normalized the velocities by the average velocity for each run. These data show that velocity falls off to about 60\% of its peak value, which occurs at the middle of the chamber, as the bead moves close to the glass surface where the apparent viscosity experienced by the bead is highest. To ensure that this effect was not due to a steric interaction with the wall, we separately measured the velocity dependence on height when the bead was moving towards the nearest surface or away from it and found identical results.

The simplest BR models predict that the velocity varies inversely with viscosity. The best fit of this model to the data of Fig.~2a, using Eq.~\ref{ViscEq} for the dependence of viscosity on bead height, yields a radius of $407 \pm 97$~nm, consistent with the size of the bead or, potentially, the actin network itself and not the actin monomers or filaments within the network, which are considerably smaller\cite{JWS:Pol2003,JWS:Cam2001}. If wall effects slow bead motion by limiting the fluctuation of individual actin filaments we would have expected to recover a much lower radius. Using Eq.~\ref{ViscEq} and our experimentally measured values for $L$ and $\eta_\infty$, we calculated the velocity as a function of viscosity (Fig. 2b).

The small separation between surfaces in these experiments relative to the size of the moving beads reduces the velocity at all positions in the chamber. Thus the average velocity computed from these experiments is predicted to be lower than if the bead were moving in free solution by a factor of 1.6, corresponding to a predicted unbounded velocity of $94 \pm 18$~nm/s. To test this prediction we measured bead motility in 80~$\mu$m--thick chambers formed by separating a coverslip from a glass slide using double-stick tape, in which wall effects are estimated to be less than 2\% on average. In these chambers, beads moved with an average velocity of $76 \pm 10$~nm/s ($n=15$), faster than in the thinner chambers.

An increase in the viscous forces acting on moving beads is too small to explain the changes in velocity with height. The total force needed to overcome drag at the measured viscosities and velocities in our experiments is on the order of 20~fN. Recent work has shown that such small forces have little effect on the velocity of actin--propelled beads, and we expect that viscous forces around 20~fN would only slow the beads by at most 0.01\%\cite{JWS:Mar2004,JWS:McG2003}, much less than the observed 36\% decrease in velocity near the walls. Instead, we propose that an increase in viscosity alters the amplitude of bead position fluctuations which impedes the growth of filaments.

Object--fluctuating BR models predict that bead velocity should scale with fluctuation amplitude. We calculated the average power spectra for bead position tracks at different distances to the surface after first removing the linear trends from the data (Fig.~3a). The amplitude of Brownian fluctuations scaled monotonically with distance to the surface over a large range of frequencies, with beads exhibiting a reduced fluctuation amplitude when closer to the surface. As expected for a BR mechanism where object fluctuations allow the intercalation of actin monomers at the tips of filaments, bead velocity increased with increasing noise amplitude (Fig.~3b). While our detrending algorithm removes much of the velocity content of the power spectra, residual variation due to the time--varying velocity limits further model--independent quantitative analysis of the power spectra (\textit{see Methods}). Specifically, because the time scales for the thermal position fluctuations and the velocity fluctuations caused by a changing number of filaments and elastic attachments overlap, it is impossible to isolate them from one another.

The nanoscale features of bead motion observed in our experiments also shed light on the role of adhesion between the growing network and bead. All BR mechanisms inherently lead to saltatory motion because of the stochastic timing of monomer addition kinetics. However, adhesion--based models, in which elastic linkages between the object and growing network attach and detach, predict an additional step--like component because the number of network linkages under stress fluctuates over time. This leads to a relaxation of the object position whenever the number of load--bearing elements changes. Both theoretical and computational adhesion--based models predict a broad distribution of nanometer--sized steps using physiologically relevant parameters\cite{JWS:Alb2004,JWS:Mog2003}. Computer simulations by Alberts and Odell show that pauses start when a number of linkages form simultaneously, and end when those linkages break together\cite{JWS:Alb2004}. Alternatively, a uniform 5.4~nm step size is predicted by models involving end-tracking motors\cite{JWS:Dic2002a} and was observed in two studies of moving \textit{Listeria} cells\cite{JWS:McG2003,JWS:Kuo2000}.

To measure the distribution of steps in our records of bead motion, we ran a pause--finding algorithm that locates positions where the magnitude of the velocity falls below a user--defined threshold value\cite{JWS:Neu2003,JWS:Alb2004} on the tracking data from both the 2--$\mu$m and 80--$\mu$m thick chambers (Fig.~4a). Histograms of the step sizes, the distance between adjacent pauses in position, and pause times, the amount of time between adjacent pauses are shown in Fig.~4b,c. For both types of chambers, the step size distribution was broad with smaller steps more probable than larger ones, in agreement with predictions from the adhesion--based models\cite{JWS:Alb2004,JWS:Mog2003} and not end--tracking models or the data from \textit{Listeria} cells.

Adhesion--based BR models also predict a correlation of the step sizes in time because pauses are induced by the rupture of a relatively small number of linkages, such that the total number does not change significantly with each step\cite{JWS:Mog2003,JWS:Alb2004}. Because the step size is largely related to the ratio of the number of breaking linkages to the total link number, many steps are thus required to significantly alter the step size. In the model of Alberts and Odell, the formation of $\sim5$ links was able to induce a pause while the total link number was approximately 100. On the other hand, if adhesion is not important, general BR models do not predict any correlation. To distinguish between these two possibilities, we performed an autocorrelation analysis on the measured step sizes (Fig.~4d). We find that the magnitude of the step sizes changes slowly with time such that the autocorrelation falls off according to a double exponential with decay constants of 17~$\pm$~1 and 383~$\pm$~7 steps, corresponding to approximately 80 and 1,800~nm of motion, respectively. This represents the distance, on average, a bead must move before the topology of the load--bearing elements in the network changes significantly.

Monomer addition kinetics have been measured to be $\sim50 s^{-1}$, considerably slower than the characteristic frequency of monomer--sized oscillations for a freely diffusing bead \cite{JWS:Pes93}. At face value these two facts seem to indicate that object fluctuations should have little effect on the overall polymerization rate. However, several factors complicate this basic picture. First, the addition of elastic connections between the object and network drastically changes the frequency distribution of any thermal motions, and therefore prediction of the timescales of bead or network fluctuations is impossible without a direct measurement of these elastic linkages. Second, thermal fluctuations also affect the rates at which the elastic connections attach and detach, and therefore may affect network growth. Current experimental data is insufficient to predict, \textit{a priori}, the effect of object fluctuations on network growth. We therefore sought to use a computational model, similar to those used by previous groups\cite{JWS:Alb2004,JWS:Mog2003}, to investigate the influence of object fluctuations in the presence of dynamic elastic connections.

We found that a minimal model, in which a fluctuating object is elastically coupled to a group of polymerizing filaments, is sufficient to predict the observed dependence of velocity on viscosity. In designing the computational model, we chose to incorporate only those features that our experiments indicated were necessary and sufficient to explain the observed phenomena, and we did not attempt to build a complete model of actin network growth. Our Monte Carlo simulations consisted of actin filaments which can grow in one dimension if sufficient space exists between the moving object and the filament end. Elastic links attach and detach to the filaments with force-dependent kinetics (additional details of modeling and simulations can be found in the Supplemental Information and Supplementary Fig.~S1).

The result of 50 ten-second simulations over a range of viscosities like those experienced by our beads correctly predicts an increase in velocity of similar magnitude to our experimental observations (Fig.~2b). Thus, even when we choose realistic polymerization kinetics and bead diffusion constants, we find a dramatic effect of the amplitude of thermal motions on the overall bead velocity. Interestingly, our simulations underestimate the velocity for very high viscosities, where bead fluctuations are highly damped. It is possible that for these situations, filament fluctuations become more important than bead oscillations and are able to drive motility.

While our data indicate that bead fluctuations influence motility, several additional factors may be present as well. First, the change in velocity with height we observe yields a size--scale of approximately 400~nm for the fluctuating element, leaving open the possibility that either the bead or the actin network is the source of the relevant fluctuations. Thermal energy must be disappated by fluctuation of the filaments and crosslinks within the network, and this breathing could have significant amplitude at length scales approaching the bead radius. We expect that future work on the rheological properties of actin networks will give insight into the amplitude of these thermal modes. Second, our data do not exclude a role for filament fluctuations in propulsion, only that under our experimental conditions they do not appear to dominate. Therefore, our data are not inconsistent with reports of immobilized beads which grow networks at the same rates as free beads\cite{JWS:Wie2003}. In this situations, we expect that network or filament fluctuations must contribute to motion production.

Taken together, these data and simulations demonstrate that actin polymerization produces mechanical work via an object--fluctuating BR mechanism, with adhesion dynamics playing a critical role in movement. Thermal fluctuations of the moving object, and not of the actin filaments, govern the velocity of actin-propelled beads, a common mimic for intracellular bacterial pathogens such as \textit{Rickettsia rickettsii}, \textit{Listeria monocytogenes} and \textit{Shigella flexneri}. Along these lines, if membrane fluctuations are found to influence motion in the case of lamellipodial and filopodial protrusions, cells could dynamically modify the tension or composition of their membranes in order to modulate the speed of protrusions and finely control actin network organization. Future research into the role of membrane tension on actin network growth will shed light on the control mechanisms used during actin--based motility.

\section{Methods}

\subsection{Motility Assays}
Carboxylated polystyrene beads 792~$\pm$~23 nm in diameter (Polysciences) were coated uniformly with the nucleation-promotion factor RickA purified from \textit{Rickettsia rickettsii} (a gift of Matthew Welch, University of California, Berkeley). Motility assays were performed by adding RickA coated beads to \textit{Xenopus laevis} egg cytoplasmic extract supplemented with an ATP-regenerating mix and Rhodamine-labeled actin for visualization of actin tails\cite{JWS:Cam99,JWS:Jen2004}. The assay mixture was immediately added to a microscope slide after preparation. For experiments in the 2~$\mu$m chambers, a low concentration of 2.1~$\pm$~0.1~$\mu$m polystyrene beads (Polysciences) were added to the mixture and a 5~$\mu$l sample was removed, squashed between a glass slide and coverslip and then sealed. For the 80~$\mu$m thick chambers, a flow chamber was made by separating a microscope slide and coverslip with two pieces of double-sided tape. Approximately 20~$\mu$l of the assay mixture was flowed through the chamber, which was then sealed.

\subsection{Three-dimensional Laser Tracking}
Our instrument is similar to one described previously\cite{JWS:Neu2003}, and it uses a single 809--nm wavelength diode laser (Blue Sky Research) to measure microsphere position using a position-sensitive detector (Pacific Silicon Sensors) that monitors scattered light in the back focal plane. A 3D, nano-positioning stage (Physik Instrumente) is used to control the position of the microscope sample relative to the optical trap with sub-nanometer precision at a bandwidth of approximately 100 Hz. In addition, a low-light camera (QImaging) is used to take fluorescence images concurrent with optical trapping. Chamber thicknesses were measured by moving a trapped bead along the microscope axis to find the contact position for both surfaces in an operation similar to the surface--finding protocol described in Lang \textit{et al.}\cite{JWS:Lan2002}

We used a software-based feedback procedure (LabView) implemented to track actin propelled beads over very long distances. In our experiments, 3D photodiode voltage signals \cite{JWS:Lan2002,JWS:Pet98,JWS:Pra99} are sampled at 2~kHz, anti-alias filtered at 1~kHz and converted into 3D position in realtime using a similar algorithm to Lang et al., but with a 5th order 3D polynomial\cite{JWS:Lan2002}. 3D calibration parameters were measured before each experiment using a bead stuck to the coverglass surface. The measured coordinates are then saved to disk and used to update the position of the microscope stage such that the microsphere remains in the center of the laser focus with a bandwidth of 100 Hz. Using this procedure, the bead remains stationary in the laboratory frame, while the surrounding fluid and actin comet tail move around it. In order to minimize external loads from the laser that might perturb actin network dynamics, we used a very low laser power to ensure that the optical force experienced by the bead was extremely small, typically less than 0.01~fN.

\subsection{Data Analysis}
To compute the velocity of each run we first calculated the time derivatives of the $x$--, $y$--  and $z$--  position versus time records by performing a 4th order Savitsky-Golay filter with a time constant of five seconds and extracting the first derivative\cite{JWS:Neu2003}. The total velocity as a function of time was then computed by adding the $x$--, $y$--, and $z$-- velocities in quadrature.

The position of a bead along its 3D pathlength, $s(t)$, can be found by integrating the magnitude of the velocity vector over the integral $[0, t]$ \cite{JWS:Wyl95}. In practice, the velocity data is significantly noisy so that it must be calculated with a reduced bandwidth. We found that 0.2~Hz was sufficient to preserve the long timescale shape of the motility paths while smoothing out the faster timescale fluctuations. Therefore, $s(t)$, as defined above, only contains information on a timescale slower than five seconds. In order to recover the pathlength at the full 2~kHz bandwidth, we integrated $\mathbf{v}(\tau)\cdot\Delta \mathbf{r}(\tau) / \vert \mathbf{v}(\tau) \vert $ over the interval $[0,t]$, where $\mathbf{v}(\tau)$ is calculated with a bandwidth of 0.2~Hz and $\mathbf{\Delta \mathbf{r}(\tau)}$ is the instantaneous change in position measured by taking the distance between adjacent points sampled at 2~kHz. This integral effectively sums only the component of the high frequency motion that lies along the average path direction.

Power spectra of bead position fluctuations were calculated using a four second analysis window. The spectra from moving beads are heavily dominated by the linear velocity component of the motion, which adds a $f^{-2}$ component at all frequencies. To examine the position fluctuations apart from this velocity contribution, we first detrended the data by subtracting the results of a linear fit from each window before performing the power spectral analysis. Because bead motion is nonuniform and the velocity changes over time, the detrending is not perfect, and the resulting power spectra decay slower than $f^{-2}$ in the high--frequency regime for both the experimental data and the simulations.

We used a pause-finding algorithm that locates times during which the magnitude of the velocity falls below a user-defined threshold value \cite{JWS:Neu2003,JWS:Alb2004,JWS:Ber2002}. Identifying pauses based the velocity magnitude allows for the detection of steps in both the positive and negative directions. For our analysis, the velocity along the trajectory was calculated using a 2nd order Savitsky-Golay filter with a time constant of 10~ms, and pauses were scored using a threshold velocity equal to one quarter of the average velocity for each run. We further reduced the effective pause-finding temporal resolution by requiring that all pauses last longer than 20~ms; dwells that were found to be shorter in duration than 20~ms were not scored.

\begin{acknowledgments}
We thank Robert Jeng and Matthew Welch for their kind gift of the RickA protein, and Sapun Parekh for help preparing the cellular extract. The authors acknowledge members of the Fletcher lab, George Oster, Chip Asbury and Sander Pronk for critical reading of the manuscript and helpful discussions. This work was supported by a Miller Institute for Basic Research in Science fellowship to J.W.S. and an NSF CAREER Award to D.A.F.
\end{acknowledgments}

\end{article}

\begin{figure}
\begin{center}
\includegraphics{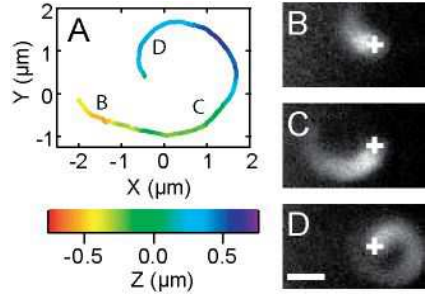}
\end{center}
\caption{(a) Three-dimensional trajectory in the laboratory frame of a single 0.8--$\mu$m bead representing about three hundred seconds of motion. The $z$--position is denoted by the color scale. (b-d) Fluorescence images of the actin tail visualized using Rhodamine-labeled actin taken at three separate time points: 12, 150 and 276 seconds respectively. Bead position for each frame along the trajectory is denoted in (a). The position of the laser focus is represented by the cross. Scale bar is 2~$\mu$m.}
\label{BRFig1}
\end{figure}

\begin{figure}
\begin{center}
\includegraphics{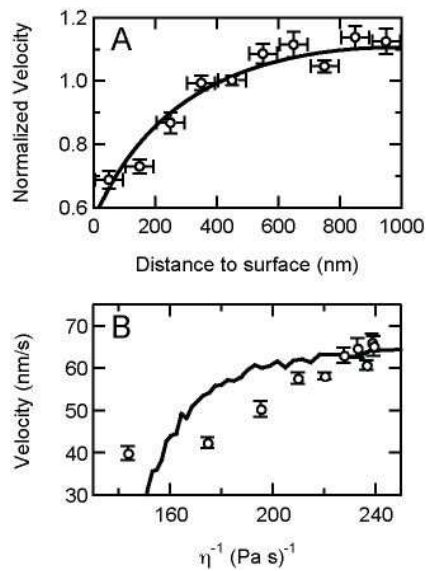}
\end{center}
\caption{(a) Normalized bead velocity as a function of distance to the nearest surface (circles) is shown for a chamber thickness of 2.5~$\mu$m. Very close to the surface the velocity is significantly lower than near the middle of the chamber. The black line represents the best fit to a model that describes the increase in viscosity due to the presence of two surfaces: $v(h)=v_\infty \left[ 1-\frac{9r}{16}\left(\frac{1}{h}+\frac{1}{L-h}\right) \right]$. Error bars in $y$ represent s.e.m. and in $x$ represent the range of distances included in each data point. (b) Velocity increases with inverse viscosity. The experimental data (circles) exhibit a similar trend to the results from simulations of an adhesion--based, object--fluctuating BR (black line).}
\label{BRFig2}
\end{figure}

\begin{figure}
\begin{center}
\includegraphics{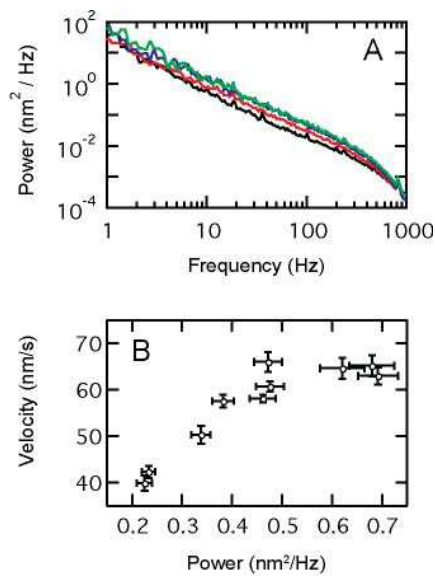}
\end{center}
\caption{(a) Average detrended power spectra of high-resolution bead motion at different positions relative to the glass surface: 50--150 nm (black), 350--450 nm (red), 650--750 nm (blue), and 950--1050 nm (green). (b) Bead velocity as a function of the Brownian noise amplitude at 20~Hz, the approximate frequency of actin monomer-sized advances at a velocity of 58~nm/s.}
\label{BRFig3}
\end{figure}

\begin{figure}
\begin{center}
\includegraphics{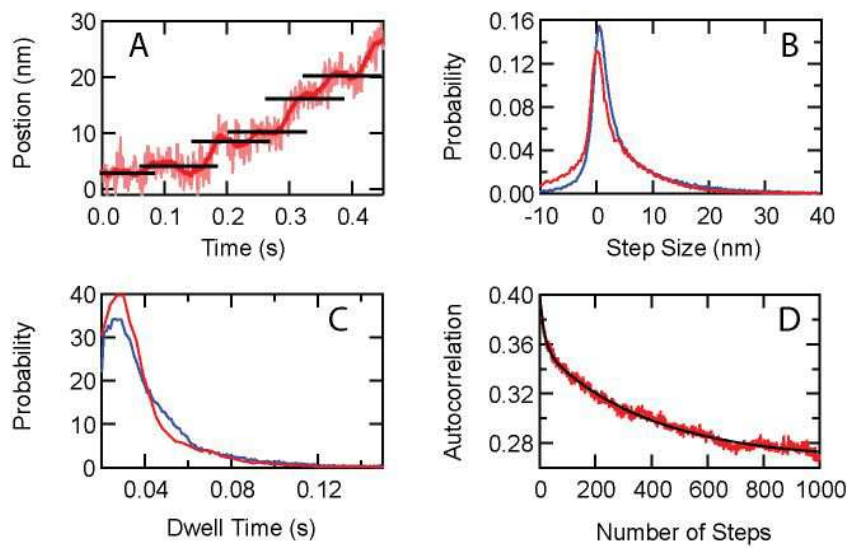}
\end{center}
\caption{(a) Sample record showing the position of one bead along its trajectory versus time. Data at a bandwidth of 2,000~Hz (pink) and smoothed to a bandwidth of 40~Hz (red) are shown. Pauses are scored by finding when the instantaneous velocity drops below a threshold value, denoted by the horizontal black lines. Step sizes are defined as the distance along the trajectory between adjacent pauses and the pause times are defined as the time between the adjacent pauses. The average step size (b) and pause time (c) distributions are both broad and show no preferred step size. Distributions were calculated from runs in both the 2--$\mu$m (blue) and 80--$\mu$m (red) chambers. (d) The autocorrelation of the measured steps sizes is averaged for all traces (red) and plotted versus the number of steps. A biphasic decay is seen with decay constants of $17 \pm 1$ and $383 \pm 7$ steps (black line).}
\label{BRFig4}
\end{figure}

\end{document}